\documentclass[preprint,aps,twocolumn,showpacs,prl,10pt]{revtex4}
\usepackage{amsfonts}
\usepackage{graphicx}
\usepackage{amsmath}
\usepackage{natbib}
\usepackage{epsfig}
\usepackage{dcolumn}

\begin{document}


\title{The origin of increase of damping
in transition metals with rare earth impurities}

\author{A. Rebei}

\email{arebei@mailaps.org}

\affiliation{Seagate Research Center, Pittsburgh, Pennsylvania
15222, USA}

\author{J. Hohlfeld}
\affiliation{Seagate Research Center, Pittsburgh, Pennsylvania
15222, USA}

\begin{abstract}
The damping due to rare earth impurities in transition metals is
discussed in the low concentration limit. \ It is shown that the
increase in damping is mainly due to the coupling of the
orbital moments of the rare earth impurities and
the conduction $p$-electrons. \ It is shown that an itinerant picture for
the host transition ions 
is needed to reproduce the observed 
dependence
of the damping on the total angular moment of the rare 
earths.
\end{abstract}


\date{April 24, 2006}


\maketitle

Magnetization dynamics has become one of the most important issues
of modern magnetism. This development is driven by the
technological demand to tailor magnetic responses on ever smaller
length and shorter time scales. The importance of this issue
manifests itself in a completely new area of research,
spintronics, and a huge literature that cannot be cited here.
Selected highlights include precessional switching by tailored
field pulses \cite{gerrits,schumacher}, spin-torque 
\cite{kiselev,kaka}, and laser-induced magnetization 
dynamics \cite{beaurepaire,hohlfeld}.

In general, magnetization dynamics is  described via the
Landau-Lifshitz-Gilbert equation (LLG) \cite{LLGB} including
additional terms to incorporate spin-torque effects \cite{zhang} or
those due to pulsed optical excitations \cite{beaurepaire2}. 
All these descriptions account for energy dissipation via a 
phenomenological damping parameter $\alpha$ which
 governs the time
needed for a non-equilibrium magnetic state to return to
equilibrium. Recently it
 has even been suggested that $\alpha$ determines
  the magnetic response to ultrafast thermal agitations
\cite{koopmans}.

Technological applications call for the ability to tailor $\alpha$ 
\cite{aps}. 
The most systematic experimental investigation on this
topic was published by Bailey {\it et al.}\,\cite{bailey1} who
studied the effect of rare-earth doping on the damping in
permalloy. Most rare earth ions induced
 a large increase of $\alpha$,
 but neither $Eu$ nor $Gd$ altered the damping of permalloy
(cf.\,Fig.\,\ref{fig001}). Since $Gd^{+++}$ and $Eu^{++}$ have
 no orbital momentum, this points
immediately to the importance of the angular momentum in the
damping process. Bailey {\it et al.\,}determined damping by 
reproducing
their data via the LLG equation using $\alpha$ as a fit parameter.
This widely used procedure points to a fundamental problem of this 
phenomenological approach. Though the LLG equation describes
 data well,
a more microscopic approach is needed to understand the origin of
damping.


  It was Elliott \cite{elliott} who first studied damping in 
semiconductors due to spin-orbit
coupling. Later Kambersky \cite{kambersky} argued that 
the Elliot-Yafet mechanism should 
be also operable in magnetic conductors. Korenman and Prange \cite{prange}
developed a more microscopic treatment and found 
that spin-orbit coupling should be important at low temperature 
in transition metals. Recent measurements of damping in 
magnetic multilayers at room temperature \cite{ingvarsson} 
suggest  that the 
{\it s-d } interaction might also be 
at the origin of damping \cite{heinrich,rebeiS}.
Howevever, all of the present models fail to 
reproduce the data of
Ref.\,\onlinecite{bailey1}.

In this Letter, we explain the increase of damping in
rare earth doped transition metals 
via a novel orbit-orbit coupling between the conduction 
electrons and the impurities. 
The well known $s$-$f$ interaction 
\cite{degennes} gives rise to a $(g_J-1)^2$ dependence of the damping 
that is in contradiction to experimental observations \cite{bailey1}.
In contrast, the orbit-orbit coupling considered here reproduces the
measured $(g_J-2)^4$ dependence of the damping. Both dependencies on the
Lande g-factor $g_J$ follow
directly from the fact that the rare earth ions are in their ground
state. Hence, their angular momentum $\mathbf{L_f}$, 
spin $\mathbf{S_f}$, and total angular momentum $\mathbf{J_f}$
 are related by the Wigner Eckard theorem:  
 $\mathbf{L_f}=(2-g_J)\mathbf{J_f}$ 
and $\mathbf{S_f}=(g_J-1)\mathbf{J_f}$. 
Deriving the magnetic moments of the transition ions from the electronic
degrees of freedom is essential 
to capture the correct behavior of damping as a function of 
$\mathbf{J}_f$.
For the uniform mode, the damping due to orbit-orbit coupling 
 is of Gilbert form in the low frequency limit.

Taking the wave functions of the
$d$-, $f$-, and conduction electrons orthogonal, the Hamiltonian 
for the rare-earth doped transition
metal in an external field $\mathbf{H}$ is
\begin{equation}
\mathcal{H}=\mathcal{H}_e + \mathcal{H}_f + \mathcal{H}_d.
\end{equation}
This approximation should be valid for the heavy
rare earths but probably fails for elements like Cerium where
valence fluctuations are 
important. The conduction electron Hamiltonian 
$\mathcal{H}_e$ is the usual one,
$ \mathcal{H}_e =\sum_{k,\sigma} \epsilon_{k,\sigma}
{a_{k,\sigma}}^\dag {a_{k,\sigma}} $, where 
${a^\dag} _{k,\sigma}$ and $a_{k,\sigma}$ are the creation and
annihilation operators of a conduction electron with momentum
$\mathbf{k}$ and spin $\sigma$. $\epsilon_{k,\sigma} $ is the
energy of the conduction electrons including a Zeeman term.

 $\mathcal{H}_f$ is the Kondo Hamiltonian \cite{kondo}
 of the localized rare earth moment
\begin{eqnarray}
\mathcal{H}_f &=&  \Gamma \mathbf{S}_e \cdot \mathbf{S}_f +\;
\lambda \mathbf{L}_e \cdot \mathbf{L}_f - \mathbf{ \mu}_f \cdot
\mathbf{H}.\label{sfHam}
\end{eqnarray}
$ \mathbf{S}_{e/f}$ and $\mathbf{L}_{e/f}$
are the spin and angular momentum of conduction and $f$ electrons,
 respectively. $\mathbf{L}_{e/f}$ are taken with respect to the 
position of the impurity.
 The spin-spin term is the well known $s$-$f$ coupling used by
 de Gennes to reproduce the Curie-temperatures in
rare earths with $\Gamma$ being
 of the order  $0.1$ eV  \cite{degennes}. 
 The last term is again a Zeeman term.  The middle 
term is the essential
orbit-orbit interaction needed in our discussion. 
To get a non-zero orbit-orbit term due to a single impurity at the 
center, it is essential to include higher terms of the partial wave 
expansion for the wave functions of the conduction electrons: 
$ \psi_k (\mathbf{r}) = \frac{4 \pi}{\sqrt{V}}
\sum_{l=0}^{\infty} \sum_{m=-l}^{m=l} i^l f(\mathbf{r})j_l(k r)
Y_{lm}(\theta_k,\phi_k)Y_{lm}^{*}(\theta,\phi) $. 
The first non-trivial contribution for $l=1$ is \cite{kondo}
\begin{equation}
\label{oo}
\mathcal{H}_{LL} =  {i} \left(2-g_J \right) \sum_{k,k^\prime}
 \lambda \left( \mathbf{k},\mathbf{k}^\prime \right)
 \hat{ \mathbf{k}}\times
\hat{\mathbf{k}}^\prime \cdot \mathbf{J}_f a_k^\dag a_{k^\prime},
\end{equation}
where the orbit-orbit coupling $\lambda$ will be assumed to be a 
function of the relative angles of the $\mathbf{k}$ vectors and is 
almost everywhere zero except for $k$ close to the Fermi 
level $k_F$. \ The magnitude of $\lambda$ is not known but 
 is expected to 
be of the same order as the spin-spin coupling constant $\Gamma$ 
\cite{levy,vanvleck}. The crystalline electric field effect in 
transition metals is less than $0.1 \; \mathrm{meV}$
which is small and
hence the spin-orbit term $\mathbf{S}_e \cdot \mathbf{L}_f$ is
neglected. \ At room temperature all the rare earth ions
 studied in Ref. \onlinecite{bailey1}
are in their ground state making the term $\mathbf{S}_f\cdot
\mathbf{L}_f$  ineffective as damping mechanism. \ This 
follows immediately from the Wigner-Eckart theorem.

\ The Hamiltonian for the
host transition ions is based on the Anderson Hamiltonian  with
explicit spin rotational invariance in the absence of a Zeeman
term \cite{anderson,prange,brazil}. It is
\begin{eqnarray}
\mathcal{H}_d & = & \epsilon_d d^{\dag}_\sigma d_\sigma + \sum_{k}
V_{kd} \left( {a^\dag}
_{k,\sigma} d_\sigma + {d^{\dag}}_\sigma a_{k,\sigma} \right) \nonumber \\
&& + \; \frac{U}{8} \rho^2 - \frac{U}{2} \mathbf{S}_d \cdot
\mathbf{S}_d   - \mathbf{\mu}_d \cdot \mathbf{H},  \label{hd}
\end{eqnarray}
where $\mathbf{S}_d$ is the
 spin operator of the local d electrons while their orbital 
angular momentum is assumed quenched. \ $\rho$ is the charge density
operator of the d electrons. In transition ions such as Ni,  
 $V_{kd} \approx 1.0-10.0 $ eV is comparable to the Coulomb potential $U$. 
\ The hybridization term between the conduction- and $d$-electrons 
 is essential  to establish a {\it spin-independent} 
orbit-orbit coupling between the $d$- and the $f$-ions. 
The degree of localization 
of the magnetic moments increases with decreasing $V_{kd}$ \cite{schrieffer}
 and controls the extent to which rare earth impurities enhance
damping.

The orbit-orbit coupling (cf.\, Eq.\,\ref{oo})  gives no
contribution for $Gd^{+++} \; (4f^7 )$ as observed 
in the experiment \cite{bailey1}. \ As
for the element $Eu$, it is believed from measurements of the
paramagnetic susceptibilities that
the ionic state is $Eu^{++} \; (4f^7)$ and not $Eu^{+++}\; (4f^6)$
\cite{degennes,thole}. \ If this is the case then clearly this is
a state with $\mathbf{L}_f = 0$ and it is the same as that of
$Gd^{+++}$. \ $Yb$ is also present in a double-ionized state 
\cite{elliottB} and therefore doping  with $Yb^{++}\; (4f^{14})$ 
should not increase damping. 
\ This result remains to be confirmed by experiment.
\ For $Eu$ there is an additional 
 reason why its angular
momentum is quenched.  The   first excited state of this 
latter element
lies only
about $ 400 \; K$ above the ground state \cite{elliottB} and
this can lift the degeneracy of the ground state. The 
average orbital angular momentum will therefore
 be zero even though $L^2$
remains a good quantum number \cite{kittelB}. \ Hence
our Hamiltonian from the outset
 reproduces the experimental results 
for $Eu$ and
$Gd$ and predicts that doping with 
 $Yb$ should not change the 
damping. \ We next address the remaining rare earth elements.

\ First, we outline the steps to derive the damping due to the
orbit-orbit coupling term. \ We are only interested in the damping
of the $d$-moments of the transition metal, therefore  it is
 advantageous
to adopt a functional integral approach. \ Since our system is
 near equilibrium  and far from
the Curie point,  we use the spin wave approximation and expand
the spin operators of the f-moments in terms of Boson operators
$f^\pm$ where  $f^{\pm}=S_{f}^{y}\pm iS_{f}^{x}$. \ We keep only
the first non-trivial terms. \  The integration of the conduction
electrons is carried out exactly. \ Afterward we integrate the
impurity variables,
 $f$ and $f^\dag$, also exactly but  keep only quartic terms in $d$ and 
$d^+$. The
remaining effective action has now only the fields $d$ and
$d^\dag$ and from their equations
 of motion  the spin propagator
$\langle m^{-}(\tau) m^{+}(\tau^\prime) \rangle$ of the $d$-moments,
$m^\pm = S_d^x \pm i S_d^y$, can be determined. We
 use a Stratonovich-Hubbard
transformation to write this effective Lagrangian in terms of
$m^\pm$. Then a stationary phase approximation of the functional
generator allows us to determine the desired propagator and hence
the damping. We finally compare the functional form of this
result to that of LLG and discuss why the electronic (itinerant)
picture of the host transition ions is essential.

\begin{figure}[hb]
  \mbox{\epsfig{file=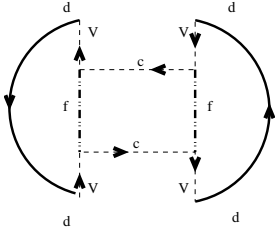,height=3 cm}}
  \caption{The first diagram that is contributing to the
damping of the d-electrons  due to the f-impurities through the
conduction electrons.} \label{fig004}
\end{figure}

\begin{figure}[ht!]
\mbox{\epsfig{file=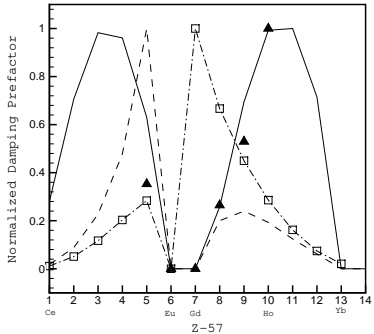,height=5 cm}}
  \caption{Comparison of the normalized leading factor 
in the damping as a function of the rare earth impurity in Eq. \ref{c} 
(solid line) and
Eq.\ref{sd} (dashed line) to the data of Ref. \onlinecite{bailey1}. The 
squares represent damping due to $s$-$f$ coupling only, Eq. \ref{sfHam}, without the orbit-orbit coupling.  }
\label{fig001}
\end{figure}

 The fundamental quantity in our calculation is  the generating
functional 
\begin{equation}
\mathcal{Z} \left[ \eta^*,\eta \right]= Tr e^{-\int_0^\beta d\tau
\{ \mathcal{H} - \eta^*(\tau) m^{-}(\tau) - \eta( \tau) m^{+}
(\tau) \}}.
\end{equation}
where $\eta$ and $\eta^*$ are external sources and $\beta$ is inverse 
temperature. The propagator, i.e.\,the connected two-point Green's function,
 of the volume mode of the transition metal ions
 is found by
functional differentiations with respect to the external sources
$\eta^*$ and $\eta$, $\langle m^{+} (\tau) m^{-} (\tau^\prime)
\rangle _c = \delta^2 ln \mathcal{Z} [ \eta*, \eta ]/\delta
\eta(\tau) \delta \eta^* (\tau^\prime)$. It is calculated within
a double random phase approximation (RPA2) method.  The {\it
true} single particle propagator of the d-bands is first found
within a RPA in the presence of an effective field due to the 
conduction electrons and the impurities.  In turn, the
effect of the $f$-impurities on the conduction electrons is
calculated within RPA. The resulting
 effective Lagrangian is now written in terms
of
 $\mathbf{m}$  only
\begin{equation}
\mathcal{L}=- \frac{1}{2} m_{ij} \mathcal{K}_{ijkl} m_{kl} - Tr
\ln\left[ G_d^{-1} + \mathcal{K} m \right].
\end{equation}
where $ {G_d}^{-1} (\sigma_1, \sigma_2) = \partial_\tau -
\bar{\epsilon}_d + V^2 G_c + tr_k \left\{ G_f G_{c} B G_{c} A
\right\} $ is the propagator of the $d$-electrons in the presence of
the conduction electrons and the rare earth impurity ($\sigma_i =
1,2$ for spin up and spin down respectively). The quadratic term
in $\mathbf{m}$ represents effective anisotropy and spin-charge
interactions and is given by
\begin{eqnarray}
\lefteqn{ \mathcal{K}_{\sigma_1 \sigma_2 \sigma_3 \sigma_4} =
\frac{-U}{4}  \left( \delta_{\sigma_1 \sigma_2} \delta_{ \sigma_3
\sigma_4} -2 \delta_{1 \sigma_1} \delta_{2 \sigma_2}
   \delta_{1 \sigma_3} \delta_{2 \sigma_4} \right)}  \\
&   - 2^2 V^4 G_f ( G_{c} B G_{c} A G_c )_{\sigma_1 \sigma_2}  G_f
( G_{c} B G_{c} A G_c )_{\sigma_3 \sigma_4} \delta_{\sigma_1
\sigma_4}   \delta_{\sigma_2 \sigma_3}
\nonumber \\
& - V^4 G_c A G_c G_f G_c B G_c.  \nonumber
\end{eqnarray}
Integrations over momentum and spin are implied in all these
expressions. The different terms that appear in $\mathcal{K}$
 are as follows: $G_{c}$ is the Green's function of the conduction
electrons in the mean field approximation
\begin{eqnarray}
\lefteqn{ G_{c}^{-1}\left(  \mathbf{k},\sigma_1,\mathbf{k}^{^{\prime}},\sigma_2
,\tau\right)   = \left(  \partial_{\tau}+\bar{\varepsilon}_{\mathbf{k}\sigma_1}%
 - \mu_F \right)  \delta_{\mathbf{kk}^{^{\prime}}}\delta_{\sigma_1\sigma_2
} } \nonumber\\
& &  + i  {\lambda ( k, k^\prime) \left(  2-g_{J}\right)  }\left\langle J_{f}%
^{z}\right\rangle \left(  k_{x}^{^{\prime}}k_{y}-k_{y}^{^{\prime}}%
k_{x}\right)  \delta_{\sigma_1\sigma_2},
\end{eqnarray}
which is off-diagonal in momentum due to the orbit-orbit coupling.
 $\bar{\varepsilon}_{k,\sigma}$ now includes Zeeman terms due to
the external field and the z-component of the field due to
impurity. The propagator $G_f$ is that of the $f$-ions in the
presence of both the conduction electrons and the transition ions,
$ G^{-1}_f\left(  \tau\right)
=\partial_{\tau}+\mu_{f}H+Tr_{k,\sigma}\left\{
G_{c}AG_{c}B\right\}$. \ The A and B matrices are  solely due to
the presence of the impurity and represent the indirect coupling
between the transition ions and the $f$-ions
\begin{eqnarray}
A\left(
\mathbf{k}^\prime,\sigma_1;\mathbf{k},\sigma_2\right)
=B\left(
\mathbf{k},\sigma_1;\mathbf{k}^\prime,\sigma_2\right)
^{\ast}=\Gamma_{0}\sigma_{\sigma_1\sigma_2}^{+}-i\lambda_{0}%
\Delta_{\mathbf{k}^{^{\prime}}\mathbf{k}}^{+}
\end{eqnarray}
where we have set  $ \Gamma_0 =
  \frac{\Gamma\sqrt{2J_{f}}}{4}  \left(  g_{J}-1\right) $,
$ \lambda_0 = \frac{\lambda\sqrt{2J_{f}}}{2 }  \left(
2-g_{J}\right)  $, and $\Delta_{k k^{\prime}} ^{\pm} = \left(
\hat{\mathbf{k}}^{{\prime}} \times \hat{\mathbf{k}}\right)
_{\pm}$.
 In the trace log term of the effective Lagrangian, 
the first nontrivial contribution  is of order $V^4$ and  
is given by 
 Fig. \ref{fig004}. The diagram with a single insertion of an
$f$-propagator does not contribute due to the antisymmetry of the
orbit-orbit coupling in the momentum space.
Varying the effective action with respect to $m_{ij}$
gives four equations which can be averaged and differentiated
with respect to the external sources  to get the $\mathbf{m}$
propagators. We are only interested in  $\mathcal{C}(1221) =
\langle m_{12} m_{21} \rangle$ which is given by
\begin{eqnarray}
\left\{G_{d11}^{-1}+ \mathcal{K}_{11ij} \langle m_{ij} \rangle
\right\} \mathcal C (1221)
+ \mathcal{K}_{11ij} \mathcal{C} (ij21)  \langle m_{12} \rangle \\
= - \langle m_{22} \rangle - \mathcal{K}_{21ij}  \mathcal{C}
(ij21)  \langle m_{22} \rangle - \mathcal{K}_{21ij} \langle m_{ij}
\rangle
 \mathcal{C} (1221)  . \nonumber
\end{eqnarray}
  In the absence of impurities, these equations are to lowest
order the time-dependent generalization of the Hartree-Fock
equations derived by Anderson \cite{anderson}. Using the RPA2 method,
we solve for
$\mathcal{C}(1221)$
\begin{eqnarray}
\lefteqn{ \mathcal{C}_{1221} (\omega_l) = \sum_{n} m_{11} (\omega_n)
m_{22} (\omega _n + \omega_l)} \\
& / \left[ 1+  \sum_{n,m} \mathcal{K}_{2112} ( \omega_m ) m_{11}
( \omega_n + \omega_m ) m_{22} 
( \omega_n + \omega_m  + \omega_l ) \right] \nonumber 
\end{eqnarray}
where $\omega_l = (2 l +1 ) \pi/\beta$ for integer $l$. If we
ignore the impurity interaction and replace the average values of
the $m_{ij}$ by the
 Anderson solution, we recover
 the RPA result for the propagator of the magnetization.  To include
the impurities, we evaluate the $d$ propagators, $m_{ij}$, within
 RPA.  In the low frequency limit, $
\omega << \Delta << \omega_c$, 
 we find that the (retarded) propagator $\mathcal{C}^R$
of the theory is proportional to $
 \left( \omega - \omega_0 + i \alpha \omega \right) ^{-1}$.
Here, $\Delta ^{-1}$ is the lifetime
of the virtual $d$ states \cite{anderson}, $\omega_c$ denotes the
frequency of the conduction electrons,
and $\omega_0$ is the ferromagnetic resonance frequency of the
transition metal.  This low
frequency limit for the damping
 is similar to that of the LLG result \cite{prange}. The
 damping $\alpha$ in the spin-conserving channel is
  proportional to $J_f (J_f + 1 ) \left( (g_J - 2) |V| \right)^4$ and is given by
\begin{eqnarray}
\lefteqn{ \alpha = c |\lambda V|^4 J_f ( J_f +1 ) (2-g_J)^4}  \label{x1}  \\
& & \times \left( \frac{ U \Delta E}{2^5 \pi^3 (E-\Delta E)^2 (E+
\Delta E)^2} \frac{( n m k_F)^2 }{18 {\omega_c}^4 } + Q(
\omega_f)\right) \nonumber \label{c}
\end{eqnarray}
 Here $n$ is the density of conduction electrons, $c$ is the
concentration of the $f$-impurities, and $E \pm \Delta E$ is the
energy of the up/down $d$ states.  These latter energies can be
determined self-consistently as in the Anderson solution
\cite{anderson} and hence their form is not expected to depend
strongly on the atomic number of the rare earth impurity at low
concentrations.  The explicit form of the function $Q$ is not
needed here but it represents contributions beyond the 'mean'
field approximation of the $f$-impurities and is given 
by Fig. \ref{fig004}.  In Fig.~\ref{fig001},
we
show that  the leading coefficient
 of the
 damping due to non-spin flip scattering (solid curve) is in very 
good agreement with the experimental 
results of Bailey  et al. \cite{bailey1}.

 Finally we point out the reasons behind insisting on using the
itinerant electrons explicitly instead of the simpler $s$-$d$ exchange
interaction which accounts well for damping in permalloy \cite{ingvarsson}.  Using a
 localized-type Hamiltonian  for the $d$-moments
\begin{eqnarray}
\mathcal{H}_d= - J \mathbf{S}_e \cdot \mathbf{S}_d
 -  \mathbf{\mu}_d \cdot \mathbf{S}_d  \label{sd}
\end{eqnarray}
instead of  Eq. \ref{hd},
leads to a damping which differs significantly from experiment (dashed curve in Fig.
\ref{fig001}).  This
localized moment Hamiltonian however appears to describe well
damping in insulators such as  heavy rare earth doped garnets 
\cite{seiden}. In garnets, the hybridization coupling is smaller than 
in metals.  Hence our result also explains why the damping in rare-earth 
doped garnets is not as strong as in the rare-earth doped transition 
metals. The experimental measurements (triangles) clearly
show that at room temperature non-spin flip scattering is more
important than spin-flip scattering which only becomes 
important close
to the critical temperature. 
 Again, the data is  well reproduced by the orbit-orbit
coupling and the relatively large increase in damping is due to
the large virtual mixing parameter $V_{kd}$.  In constrast, the
  $s$-$f$ coupling (squares in Fig. \ref{fig001}) 
is  in conflict with experiment.

 In summary, we have shown that the damping in rare-earth doped
transition metals is mainly due to an orbit-orbit coupling between
the conduction electrons and the impurity ions.  For near
equilibrium conditions and in the 
low frequency regime this leads to damping for the uniform mode
that is of Gilbert form.
 The orbit-orbit mechanism introduced here is 
 much stronger than
the spin-orbit based Elliott-Yafet-Kambersky mechanism since the
charge-spin coupling at the host ion is of the order of 1-10 eV
compared to 0.01 eV for spin-orbit coupling.  The predicted
increase of damping is proportional to $V^4$ which in transition
ions is of the same order as $U$ the Coulomb potential.  A
localized model for the d-moments based on the $s$-$d$ exchange is
unable to account for the  increase in damping in these doped
systems as a function of the orbital moment 
of the rare-earth impurities.   An additional 
test of this damping theory would be to 
measure the effect of a single rare earth element on the damping 
in various transition metals.  Such experiments will provide 
further insight into the dependence of damping on $V$ 
and will improve our 
understanding of the itinerant versus localized pictures
 of magnetism.

\ We acknowledge fruitful discussions
with P. Asselin, O. Heinonen, P. Jones, O. Myarosov,
and Y. Tserkovnyak.

\end{document}